\documentclass[aps,pra,twocolumn,showpacs,floatfix]{revtex4}
\usepackage{graphicx}
\usepackage{nicefrac}
\usepackage{amsmath}
\usepackage{amsfonts}
\usepackage{amssymb}
\usepackage{amsthm}
\usepackage{epsf}
\usepackage{bm}
\usepackage{bbm}
\usepackage{longtable}

\usepackage{dcolumn}

\newcolumntype{.}{D{x}{}{-1}}
\newcommand{\lbr}{\langle}
\newcommand{\rbr}{\rangle}

\newcommand{\Za}{{Z \alpha}}

\newcommand{\vare}{\varepsilon}

\newcommand{\pr}{^{\prime}}
\newcommand{\ppr}{^{\prime\prime}}
\newcommand{\pppr}{^{\prime\prime\prime}}

\newcommand{\cross}[1]{#1\!\!\!/}

\begin{document}

\title{Electron self-energy in the presence of magnetic field:\\
hyperfine splitting and $\boldsymbol{g}$ factor}

\author{Vladimir A. Yerokhin}
\affiliation{Max--Planck--Institut f\"ur Kernphysik,
Postfach 10 39 80, 69029 Heidelberg, Germany}
\affiliation{Center
for Advanced Studies, St.~Petersburg State Polytechnical
University, Polytekhnicheskaya 29, St.~Petersburg 195251, Russia}

\author{Ulrich D. Jentschura}
\affiliation{Max--Planck--Institut f\"ur Kernphysik,
Postfach 10 39 80, 69029 Heidelberg, Germany}
\affiliation{Institut f\"ur Theoretische Physik, Universit\"at Heidelberg,
Philosophenweg 16, 69120 Heidelberg, Germany}

\begin{abstract}

A high-precision numerical calculation is reported for the self-energy
correction to the hyperfine splitting and to the bound-electron $g$ factor in
hydrogenlike ions with low nuclear charge numbers. The binding nuclear Coulomb
field is treated to all orders, and the nonperturbative remainder beyond the
known $Z\alpha$-expansion coefficients is determined. For the $^3{\rm He}^+$
ion, the nonperturbative remainder yields a contribution of $-$450~Hz to the
normalized difference of the $1S$ and $2S$ hyperfine-structure intervals, to be
compared with the experimental uncertainty of 71~Hz and with the theoretical
error of 50~Hz due to other contributions. In the case of the $g$ factor, the
calculation provides the most stringent test of equivalence of the perturbative
and nonperturbative approaches reported so far in the bound-state QED
calculations.

\end{abstract}


\maketitle

%
%
The hyperfine structure (hfs) of the ground state of hydrogen is experimentally
known with a relative accuracy of $1\times 10^{-12}$ \cite{essen:71}, this
measurement having for a long time been among the most precise ones in physics.
One of the remarkable features of the hfs is an important role of the binding
effects in its theoretical description. For the self-energy (SE) correction to the
hfs, the binding effects change the sign of the correction already for a nuclear
charge number $Z=8$ and make the expansion in the binding-strength parameter $\Za$
completely inadequate for high values of $Z$ (here, $\alpha$ is the fine-structure
constant). Large coefficients of the $\Za$ expansion and the high accuracy of
experimental results call for an all-order (in $\Za$) approach in the theoretical
description of the hfs even for systems as light as hydrogen.

The high-precision all-order calculation of radiative corrections for hydrogen is
a notoriously difficult problem. This point can be illustrated by considering the
SE correction to the Lamb shift. Its accurate evaluation to all orders in $\Za$
was first accomplished by P.~J.~Mohr in 1974 \cite{mohr:74:a} for $Z\ge 10$, while
an analogous calculation for $Z=1$ was not realized until two decades later
\cite{jentschura:99:prl}.

All-order calculations of the SE correction to the hfs started in late 1990s
\cite{persson:96:prl,shabaev:96:pisma,blundell:97:pra}. The first attempt at a
numerical evaluation for hydrogen was made at the same time in
Ref.~\cite{blundell:97:prl}. Due to insufficient numerical accuracy at $Z=1$ in
that work, the goal was reached in an indirect way: the known terms of the $\Za$
expansion were subtracted from the all-order numerical results for $Z \ge 5$, and
the higher-order remainder thus inferred was extrapolated towards $Z=1$. The
result obtained was used as an important theoretical input for the determination
of the muon mass from the muonium hfs measurements \cite{mohr:05:rmp}.

The accuracy of the numerical evaluation of the SE correction to the hfs was
improved by several orders of magnitude during the past years
\cite{yerokhin:01:hfs,yerokhin:05:hfs}. However, the precision obtained was
still insufficient for a direct determination of the higher-order SE remainder
at $Z=1$, and an extrapolation procedure was again employed. The studies
\cite{yerokhin:01:hfs,yerokhin:05:hfs} provided a remainder value for the
normalized difference of the $1S$ and $2S$ hfs intervals, $\Delta_2 = 8\Delta
E_{2S}-\Delta E_{1S}$ \cite{karshenboim:02:epjd}, in $^3{\rm He}^+$ and
demonstrated a $2\sigma$ deviation of the theoretical prediction from the
experimental result \cite{schluesser:69,prior:77}. The accuracy of the
extrapolation procedure of Refs.~\cite{yerokhin:01:hfs,yerokhin:05:hfs} has
recently become a subject of some concern. In particular, there is an opinion
\cite{karshenboim:05:cjp} that the error of the extrapolation is four times
larger than given in Refs.~\cite{yerokhin:01:hfs,yerokhin:05:hfs}, which would
bring theory and experiment back into agreement.

The main goal of the present investigation is to perform the first direct,
high-precision theoretical determination of the higher-order remainder of the
SE correction to the hfs of the $1S$ and $2S$ states of light hydrogenlike
ions. In addition, we carry out a related study of the SE correction in the
presence of an external homogeneous magnetic field, i.e., the SE correction to
the $g$ factor of an electron bound by a spinless nucleus.

High-precision experimental investigations for the bound-electron $g$ factor have
a shorter history than those for the hfs but are not less important. A relative
accuracy of $5\times10^{-10}$ was reached in recent microwave measurements in
hydrogenlike carbon and oxygen \cite{haeffner:00:prl,verdu:04}, thus providing a
new tool for the determination of the electron mass \cite{remark}. A recent
proposal \cite{quint:submitted} to employ laser spectroscopic techniques in these
measurements opens perspectives for improving the experimental accuracy
(particularly, for the helium ion) up to the level of $10^{-12}$.

Already at the present level of experimental accuracy, the theoretical description
of the bound-electron $g$ factor has to be performed to all orders in $\Za$. The
numerical precision often becomes a matter of crucial importance in such
calculations. So, an increase of the numerical accuracy for the SE correction to
the $g$ factor by an order of magnitude achieved in Ref.~\cite{yerokhin:02:prl} as
compared to the previous evaluations
\cite{persson:97:g,blundell:97:pra,beier:00:pra} resulted in an improvement in the
electron mass value. In order to match the $10^{-12}$ level of accuracy
anticipated in future experiments on the helium ion, the precision of numerical
calculations of the SE correction should be enhanced by several orders of
magnitude. This task will be accomplished in the present work.

\begin{figure}
\centerline{
\resizebox{0.9\columnwidth}{!}{%
  \includegraphics{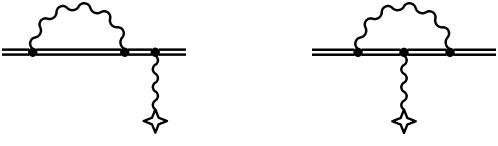}
}}
 \caption{ The electron self-energy in the presence of a magnetic field. The double
line indicates the bound electron and the wave line with a cross is the magnetic
field. \label{fig:segfact}}
\end{figure}

The SE correction in the presence of an external (magnetic) potential $V_{\rm
magn}$ is graphically represented by two topologically nonequivalent diagrams in
Fig.~\ref{fig:segfact}. Formal expressions for them can be obtained by considering
a first-order perturbation of the SE correction to the Lamb shift by $V_{\rm
magn}$. Perturbations of the reference-state wave function, the binding energy,
and the electron propagator give rise to the irreducible, the reducible, and the
vertex contributions, respectively. General formulas for these contributions are
known and can be found in our previous study \cite{yerokhin:05:hfs}; for a
detailed analysis we direct the reader to Ref.~\cite{sunnergren:98}. The
irreducible part reads
\begin{equation}
\Delta E_{\rm ir} = \lbr a| \Sigma(\vare_a)|\delta a\rbr
                 +  \lbr \delta a| \Sigma(\vare_a)|a\rbr\,,
\end{equation}
where $\Sigma(\vare_a)$ is the SE operator defined so that its diagonal matrix
element $\lbr a| \Sigma(\vare_a)|a\rbr$ yields the one-loop SE correction to the
Lamb shift \cite{mohr:74:a}, and $|\delta a\rbr$ is the first-order perturbation
of the reference-state wave function $|a\rbr$ by $V_{\rm magn}$. The reducible
part is given by
\begin{equation}
\Delta E_{\rm red} = \lbr a|V_{\rm magn}|a\rbr\,
        \left.     \lbr a| \frac{\partial}{\partial \vare}
         \Sigma(\vare)
           |a\rbr  \right|_{\vare=\vare_a} \,,
\end{equation}
and the vertex part is
\begin{align}
\Delta E_{\rm ver} &\ = \frac{ie^2}{2\pi}\int_{-\infty}^{\infty}d\omega\,
\sum_{n_1n_2} \nonumber \\ & \times \frac{\lbr n_1|V_{\rm magn}|n_2\rbr\, \lbr
an_2|\alpha_{\mu}\alpha_{\nu}D^{\mu\nu}(\omega)|n_1a\rbr}
{(\vare_a-\omega-u\vare_{n_1})(\vare_a-\omega-u\vare_{n_2})}\,,
\end{align}
where $\alpha_{\mu} = (1,\bm{\alpha})$ are the Dirac matrices,  $u = 1-i0$, and
$D^{\mu\nu}$ is the photon propagator.

The calculation of the irreducible part is similar to the evaluation of the
diagonal matrix element of the SE operator. It is performed here by a
generalization of the approach of Ref.~\cite{jentschura:99:prl}, with the use
of the closed-form analytical representation for the perturbed wave function
$|\delta a\rbr$ \cite{shabaev:91:jpb}. The evaluation of the reducible and
vertex parts is more difficult. It is carried out after splitting them into
several parts,
\begin{eqnarray}
\Delta E_{\rm red} &=&  \Delta E^{(a)}_{\rm red} + \Delta E^{(0)}_{\rm red}
   + \Delta E^{(1+)}_{\rm red} \,,
   \\
\Delta E_{\rm ver} &=& \Delta E^{(a)}_{\rm ver} + \Delta E^{(0)}_{\rm ver}
   + \Delta E^{(1)}_{\rm ver}+ \Delta E^{(2+)}_{\rm ver} \,,
\end{eqnarray}
where the upper index $(a)$ labels the contributions induced by the
reference-state part of the electron propagators and the other indices specify the
total number of interactions with the binding field in the electron propagators
[the index $(i+)$ labels the terms generated by $\ge$$i$ such interactions]. The
reference-state contributions $\Delta E^{(a)}_{\rm red}$ and  $\Delta E^{(a)}_{\rm
ver}$ are separately infrared divergent. The divergences disappear when the
contributions are regularized in the same way and evaluated together. The
zero-potential parts $\Delta E^{(0)}_{\rm red}$ and $\Delta E^{(0)}_{\rm ver}$ are
separately ultraviolet divergent. They are covariantly regularized by working in
an extended number of dimensions and calculated in momentum space. The remainder
of the reducible part $\Delta E^{(1+)}_{\rm red}$ contains at least one
interaction with the binding field in the electron propagators and is finite. In
its evaluation, advantage was taken of a generalization of the numerical
procedures originally developed in Ref.~\cite{jentschura:99:prl}.

The remaining vertex contributions $\Delta E^{(1)}_{\rm ver}$ and $\Delta
E^{(2+)}_{\rm ver}$ contain three electron propagators and represent the most
difficult part of the calculation. The key to the success was to isolate the
one-potential vertex contribution  $\Delta E^{(1)}_{\rm ver}$ and to calculate it
without any partial-wave expansion in momentum space. For the SE correction to the
$g$ factor, such a calculation has been carried out in
Ref.~\cite{yerokhin:02:prl}, employing the fortunate fact that in momentum space,
the interaction with the homogeneous magnetic field is expressed in terms of the
(gradient of the) $\delta$-function. This is not the case for the hfs, and the
calculation of this contribution is much more difficult. The general expression
for $\Delta E^{(1)}_{\rm ver}$ reads
\begin{align} \label{ee1}
\Delta E^{(1)}_{\rm ver}  = &\  -8\pi i\alpha \int \frac{d\bm{p} \, d \bm{p}\pr
\, d\bm{p}\ppr}{(2\pi)^9} \int \frac{d^4k}{(2\pi)^4}\,
\frac{V_C(\bm{p}\ppr)}{k^2} \nonumber \\ &\ \times \overline{\psi}_a(\bm{p})\,
\gamma_{\sigma} S(p-k)\, \gamma_0 S(p-p\ppr-k)\, \nonumber \\ &\ \times
\gamma_0V_{\rm magn}(\bm{p}\pppr)\, S(p\pr-k)\,\gamma^{\sigma}\,
\psi_a(\bm{p}\pr)\,,
\end{align}
where $p\pppr = p-p\pr-p\ppr$, $p_0 = p\pr_0 = \vare_a$, $p_0\ppr =p_0\pppr = 0$,
$S(p) = 1/(\cross{p}-m)$ is the free electron propagator, and $V_C$ is the Coulomb
potential. Effectively, $\Delta E^{(1)}_{\rm ver}$ is a two-loop contribution
because two momentum integrations (over $d^4k$ and $d\bm{p}\ppr$) need to be
performed analytically. They are carried out after joining denominators by
introducing 4 auxiliary Feynman parameters. Next, we integrate over all angular
variables except $\bm{p}\cdot\bm{p}\pr$, which leaves 7 integrations (3 over the
kinematic variables and 4 over the auxiliary parameters) to be carried out
numerically. The numerical evaluation is rather time-consuming (about a month of
processor time for each value of $Z$) but the crucial point is that it does not
involve any partial wave summations.

The remaining vertex contribution $\Delta E^{(2+)}_{\rm ver}$ contains bound
electron propagators, and so the partial-wave expansion in its evaluation is
unavoidable. However, the convergence of this expansion turns out to be very good,
provided that the integrations over all radial variables are first carried out.
For instance, at $Z=1$, the sum of only the first two partial waves for the hfs
yields a result with a relative accuracy of $10^{-5}$. The good convergence is due
to the separation of the one-potential contribution $\Delta E^{(1)}_{\rm ver}$
introduced in this work. About 120 partial waves included in the actual
calculation and the extended-precision arithmetics employed allowed us to control
the calculation to a level of $10^{-9}$.

The results for the SE correction to the hfs can be conveniently represented as
\begin{align} \label{ee2}
  \Delta E_{n} &\ = E_F(n)\, \frac{\alpha}{\pi}\,
    \Bigl[a_{00}+ (\Za)\,a_{10}
+ (\Za)^2\bigl(
         L^2 a_{22}+ L\,a_{21}
   \nonumber \\ &\
+a_{20}\bigr)
     + (\Za)^3\,L\,a_{31}+ (\Za)^3\,F_{n}(\Za)
         \Bigr]\,,
\end{align}
where $E_F(n)$ is the non-relativistic hfs value, $L =\ln[(Z\alpha)^{-2}]$, and
$a_{ij}$ are coefficients of the $Z\alpha$-expansion known today: $a_{00}(nS) =
1/2$, $a_{10}(nS) = -8.03259003$, $a_{22}(nS) =-2/3$, $a_{21}(1S) = -1.334504$,
$a_{21}(2S) = 0.317104$, $a_{20}(1S) = 17.122339$, $a_{20}(2S) = 11.901105$,
$a_{31}(nS) = -13.307416$, see recent works
\cite{sehfs:expansion,karshenboim:02:epjd} and references therein for earlier
studies. $F_{n}$ is the higher-order remainder, which should addressed in a
numerical all-order approach.

%
%
\begin{table}
\caption{SE correction to the hfs of $nS$ states of hydrogenlike ions. $\delta E_n
= \Delta E_n/[(\alpha/\pi)\, E_F(n)]$ and $F_n$ is defined in Eq.~(\ref{ee2}).
 \label{tab:hfs} }
\begin{ruledtabular}
\begin{tabular}{cc..c}
$n$ & $Z$ &  \multicolumn{1}{c}{$\delta E_n$} & \multicolumn{1}{c}{$F_n$} & Ref. \\
\hline\\[-9pt]
 1 & 1 & 0.438\,10x1\,842\,(2)  & -13.8x308\,(42) \\
   &   &                        & -13.8x\,(3)       & \cite{yerokhin:05:hfs} \\
   &   &                        & -15.9x\,(1.6)      &  \cite{nio:01} \\
   &   &                        & -12\,\,(x2)         & \cite{blundell:97:prl} \\
   & 2 & 0.373\,46x7\,603\,(3)  & -14.1x159\,(10) \\
   & 3 & 0.307\,58x3\,837\,(4)  & -14.4x120\,(4) \\
   & 4 & 0.241\,00x5\,729\,(6)  & -14.6x962\,(2) \\
   & 5 & 0.174\,02x6\,210\,(7)  & -14.9x673\,(2) \\
\hline\\[-9pt]
 2 & 1 & 0.438\,69x2\,275\,(3)  &  -6.1x205\,(84) \\
   &   &                        &  -6.2x\,(9)        & \cite{yerokhin:05:hfs} \\
   &   &                        &  -7.8x\,(1.4)      & \cite{yerokhin:01:hfs} \\
   & 2 & 0.375\,35x2\,040\,(4)  &  -6.9x129\,(11) \\
   &   &                        &  -6.9x\,(4)        & \cite{yerokhin:05:hfs} \\
   &   &                        &  -8.2x\,(9)        & \cite{yerokhin:01:hfs} \\
   & 3 & 0.311\,20x3\,192\,(5)  &  -7.5x833\,(5) \\
   & 4 & 0.246\,66x5\,422\,(7)  &  -8.1x698\,(3) \\
   & 5 & 0.181\,93x8\,683\,(10) &  -8.7x069\,(2) \\
\end{tabular}
\end{ruledtabular}
\end{table}
%
%
\begin{table}
\caption{SE correction to the $1S$ bound-electron $g$ factor, in units of
$10^{-6}$ (ppm).  $H_1$ is the higher-order remainder defined by Eq.~(\ref{ee3})
and obtained by a direct evaluation; $H_1(\rm extr.)$ denotes the extrapolated
results. The results of Refs.~\cite{beier:00:pra,yerokhin:02:prl} are scaled for
the present value of $\alpha$ \cite{mohr:05:rmp}.
 \label{tab:gfact} }
\begin{ruledtabular}
\begin{tabular}{c...c}
    $Z$    &  \multicolumn{1}{c}{$\Delta g_1$}
                             &   \multicolumn{1}{c}{$H_1$}
                       & \multicolumn{1}{c}{$H_1$(\rm extr.)} & Ref. \\
\hline\\[-9pt]
 1 & 2\,322.8x40\,245\,(1) &   12\,(x31)   & 23.3x9\,(80) & \\
   & 2\,322.8x40\,3\,(1)   &               &       &  \cite{yerokhin:02:prl} \\
   & 2\,322.8x40\,2\,(9)   &               &       &  \cite{beier:00:pra} \\
 2 & 2\,322.9x04\,052\,(4) &   23.1x(2.8)  & 23.0x3\,(44) & \\
 3 & 2\,323.0x14\,310\,(8) &   22.8x8(70)  &       & \\
 4 & 2\,323.1x75\,54\,(2)  &   22.5x7(30)  &       & \\
 5 & 2\,323.3x92\,99\,(2)  &   22.3x5(16)  &       & \\
\end{tabular}
\end{ruledtabular}
\end{table}

The results of our numerical calculation of the SE correction to the hfs of the
$1S$ and $2S$ states of light hydrogenlike ions with $Z\le 5$ are presented in
Table~\ref{tab:hfs}. The fine-structure constant of $\alpha^{-1} = 137.03599911$
\cite{mohr:05:rmp} was employed in the calculation. Since the current uncertainty
of $\alpha$ (3 ppb) does not influence the numerical accuracy of the higher-order
remainders, $\alpha$ is assumed to attain exactly the value indicated. Good
agreement is observed with the extrapolated values of the higher-order remainder
obtained previously \cite{blundell:97:prl,yerokhin:01:hfs,yerokhin:05:hfs} and
with the $\Za$-expansion result of Ref.~\cite{nio:01}, but their accuracy is
increased by several orders of magnitude.

Our calculation removes a significant source of uncertainty in the theoretical
predictions for the normalized difference of the $1S$ and $2S$ hfs intervals in
hydrogen and helium-3 ion, $\Delta_2 = 8\Delta E_{2}-\Delta E_{1}$. For
$^3$He$^+$, the SE remainder determined in this work amounts to $-0.450$~kHz.
Combining this result with other theoretical contributions to $\Delta_2$ described
in detail in
Refs.~\cite{karshenboim:02:epjd,jentschura:06:hfs,karshenboim:05:report}, we
obtain the total theoretical value $\Delta_2(^3{\rm He}^+)_{\rm theo} =
-1190.135\,(50)$~kHz, to be compared with the experimental result $\Delta_2(^3{\rm
He}^+)_{\rm exp} = -1189.979\,(71)$~kHz \cite{schluesser:69,prior:77}. Our
calculation of the SE remainder improves the accuracy of the theoretical
prediction by a factor of three, as compared with
Ref.~\cite{karshenboim:05:report}. For hydrogen, the theoretical and experimental
results read $\Delta_2({\rm H})_{\rm theo} = 48.9541(23)$~kHz and $\Delta_2({\rm
H})_{\rm exp} = 49.13\,(13)$~kHz \cite{kolachevsky:04:prl,essen:71},
correspondingly.

For the $g$ factor, the results of our numerical evaluation can be
parameterized as
\begin{align} \label{ee3}
  \Delta g_{n}  = \frac{\alpha}{\pi}\,
    \Bigl[1+  (\Za)^2\,b_{20} &\
+ (\Za)^4\bigl(L\,b_{41}
+b_{40}\bigr)
   \nonumber \\ &\
     + (\Za)^5\,H_{n}(\Za)
         \Bigr]\,,
\end{align}
where $b_{ij}$ are known coefficients of the $\Za$ expansion: $b_{20}(nS) =
\frac16\,n^{-2}$, $b_{41}(nS) = \frac{32}{9}\, n^{-3}$, $b_{40}(1S) = -10.236524$,
$b_{40}(2S) = -1.338464$, see Ref.~\cite{pachucki:04:prl:gfact} and references
therein. $H_n$ is the remainder incorporating all higher-order contributions. It
is remarkable that for the $g$ factor, the higher-order remainder enters in the
relative order $(\Za)^5$ rather than in the relative order $(\Za)^3$, as in the
case of the hfs. This means that cancellations in extracting the remainder from
numerical results for $Z=1$ are by four orders of magnitude larger for the $g$
factor than for the hfs.

The results of our numerical calculation of the SE correction for the $1S$
bound-electron $g$ factor are presented in Table~\ref{tab:gfact}. For hydrogen,
they are consistent with values reported previously
\cite{beier:00:pra,yerokhin:02:prl} but are by two orders of magnitude more
accurate. At the same time, all ten digits of our numerical all-order result for
$Z=1$ coincide with the value obtained within the $\Za$-expansion. The fact of
this coincidence can be considered as one of the most stringent tests of
consistency of the two main theoretical approaches presently developed in
bound-state QED. Similar agreement between the $\Za$-expansion and the all-order
approach was observed for the $2S$ state; the corresponding results will be
presented elsewhere.

The accuracy of the direct numerical determination of the $g$-factor remainder
$H_1$ for $Z=1$ and $2$ can be increased by extrapolating our results obtained for
the higher-$Z$ region. We employ the extrapolation procedure described in
Refs.~\cite{yerokhin:01:hfs,yerokhin:05:hfs} and the numerical data for the
remainder $H_1$ for $Z$ as high as 20 in order to obtain the improved results
listed in Table~\ref{tab:gfact} under the label $H_1(\rm extr.)$.

To conclude, we have performed high-precision all-order calculations for the SE
correction to the hfs and to the bound-electron $g$ factor in light hydrogenlike
systems, improving the numerical accuracy by several orders of magnitude as
compared to the previous evaluations. Accurate nonperturbative results have been
obtained for the higher-order SE remainder for the hfs. We remove an important
source of uncertainty in theoretical predictions for the normalized difference of
the $1S$ and $2S$ hfs intervals in hydrogen and the helium-3 ion and increase the
theoretical accuracy by a factor of three.

Valuable discussions with A.~I.~Milstein are gratefully acknowledged. The work
was supported by DFG (grant No.~436 RUS 113/853/0-1). V.A.Y. acknowledges
support from RFBR (grant No.~06-02-04007) and the foundation ``Dynasty.''
U.D.J. acknowledges support from DFG (Heisenberg program).


\begin{thebibliography}{10}
\bibitem{remark}
As a result, the latest values of the mass of the electron and of the ratio of the
electron to proton mass are derived mainly from these experiments (see
Ref.~\cite{mohr:05:rmp}).
\end{thebibliography}
\end{document}